\documentclass[twocolumn,pra,showpacs,superscriptaddress,amssymb,amsmath]{revtex4}
\usepackage{graphicx}
\usepackage{epstopdf}
\usepackage{hyperref}

\hypersetup{%
  pdfpagemode=None, 
  pdfstartpage=1,
  pdfstartview=FitH,
  pdftoolbar=true,
  colorlinks = true,
  linkcolor=blue,
  citecolor=blue,
  bookmarksopen=false
}

\begin{document}

\title{Asymptotic model for shape resonance control of diatomics\\
  by intense non-resonant light}

\author{Anne Crubellier}
\email{anne.crubellier@u-psud.fr}
\affiliation{Laboratoire Aim\'e Cotton, CNRS, Universit\'e Paris-Sud 11, 
  ENS Cachan, B\^atiment 505, 91405  Orsay Cedex, France} 

\author{Rosario Gonz\'alez-F\'erez}
\email{rogonzal@ugr.es}
\affiliation{Instituto Carlos I de F\'isica
    Te\'orica y Computacional and Departamento de F\'isica At\'omica,
    Molecular y Nuclear, Universidad de Granada, 18071 Granada,
    Spain}
\affiliation{The Hamburg Center for Ultrafast Imaging, University of Hamburg,
      Luruper Chaussee 149, 22761 Hamburg, Germany}
\author{Christiane P. Koch}
\email{christiane.koch@uni-kassel.de}
\affiliation{Theoretische Physik,  Universit\"at Kassel,
  Heinrich-Plett-Str. 40, 34132 Kassel, Germany}

\author{Eliane Luc-Koenig}
\email{eliane.luc@u-psud.fr}
\affiliation{Laboratoire Aim\'e Cotton, CNRS, Universit\'e Paris-Sud 11, 
  ENS Cachan, B\^atiment 505, 91405  Orsay Cedex, France}

\date{\today}
\begin{abstract}
  We derive a universal model for atom pairs interacting with
  non-resonant light via the polarizability anisotropy, based on the
  long range properties of the   scattering. 
  The corresponding dynamics can be obtained 
  using a nodal line technique to
  solve the asymptotic Schr\"odinger equation. It consists in 
  imposing physical boundary conditions at long range
  and vanishing of  the wavefunction  at a
  position separating inner zone and asymptotic region. We show that
  nodal lines which depend on the intensity of the non-resonant light
  can satisfactorily account for the effect of the polarizability at
  short range.   The approach
  allows to determine the resonance structure, energy, width, 
  channel mixing and hybridization even for narrow resonances.
\end{abstract}
\maketitle

\section{Introduction}
\label{sec:intro}
Ultracold collisions have been a focus of AMO physics research for the
last two decades. The keen interest in the subject is due to two main
aspects -- collisions at very low energy are highly non-classical, and
they show universal behavior~\cite{Weiner98,BurnettNatureReview02}. 
The quantum nature of ultracold collisions implies that the dynamics
are governed by tunneling and resonances. The latter are at the core
of an unprecedented control over the scattering particles that was 
achieved experimentally~\cite{ChinRMP10}. At the same time, the
universal behavior of ultracold collisions has given 
rise to a thourough understanding of the underlying 
dynamics. For example, quantum-defect theory can be employed to
calculate atom-atom scattering properties and bound 
rovibrational levels close to
threshold~\cite{GaoPRA98,GaoPRA01,GaoJPB03}. 

A theory based solely on the asymptotic properties of the interaction
potential has proven useful also for the description of 
photoassociation~\cite{CrubellierJPB06}, i.e., the light-assisted
formation of molecules~\cite{Weiner98,JonesRMP06}.
In particular, the nodal line technique to solve the Schr\"odinger
equation in the asymptotic approximation was employed to the determine
the scattering length~\cite{CrubellierEPJD99,PasquiouPRA10}
and potential energy curves~\cite{VanhaeckeEPJD04} in several diatomic
molecules. The formalism was extended to shape
resonances~\cite{GaoPRA09,LondonoPRA10}, which occur when 
a scattering state becomes trapped behind the centrifugal barrier for
partial waves with $\ell >0$. This extension has allowed 
to capture all essentials of shape resonances in terms of
a single parameter, the $s$-wave scattering length which universally
characterizes the long-range two-body interaction. 

An important aspect of shape resonances is that they lead
to an increased pair density at short interatomic 
separations~\cite{KochJPB06} and are thus crucial for molecule
formation at ultralow temperatures~\cite{ChinRMP10,BoestenPRL96}.
However, due to the rotational
excitation involved in generating the centrifugal barrier, the lowest
energies at which shape resonances occur typically correspond to
temperatures of a few milli-Kelvin. The
interaction of  non-resonant light with the polarizability of the
atom pair can be used to shift the positions of shape
resonances to lower energies~\cite{Aganoglu11,GonzalezPRA12}.
If the resonance position is made to match the trap temperature,
the photoassociation rates are predicted to go up by two to three
orders of magnitude~\cite{GonzalezPRA12}. This control
is of a universal character, independent of the frequency  of the
light  and the  energy level structure of the molecule  (as long  as
the frequency remains far from any molecular resonance).
Interestingly, non-resonant light control 
should also enable magnetoassociation by creating new Feshbach
resonances and by strongly enlarging their
width~\cite{TomzaPRL14}. 
Non-resonant field also affects bound rovibrational
levels by shifting their energies and  hybridizing  their rotational 
motion~\cite{GonzalezPRA12,TomzaMolPhys13}. This leads to 
alignment of the wave functions along the field
direction~\cite{FriedrichPRL95}. 

These manifold proposals for control using non-resonant light call for an
extension of asymptotic models~\cite{GaoPRA09,LondonoPRA10} to account
for the coupling with non-resonant light via the polarizability
anisotropy. Such an approach 
is promising as long as the relevant physics occurs at large
interatomic separations and in an energy region close to threshold. 
This is the case both for shape resonance control
in photoassociation~\cite{GonzalezPRA12} or Feshbach resonance
engineering~\cite{TomzaPRL14}. The dependence of the polarizability
on interatomic separation is then universal and depends
only on the polarizabilities of the constituent
atoms~\cite{HeijmenMP96,JensenJCP02}. 
Including the interaction with a non-resonant field in asymptotic
models should allow for predicting the field intensity
that is required to modify the position of a shape resonance by a
desired amount without exact knowledge of the potential. This is the
question that we address here. 

We test the asymptotic model against exact results for the strontium
dimer which has recently been the subject of intense research
both experimentally~\cite{SteinPRA08,EscobarPRA08,StellmerPRL12,ReinaudiPRL12} 
and theoretically~\cite{SkomorowskiKochPRA12,SkomorowskiKochJCP12}. The
interest in Sr$_2$ is motivated by prospects to study the variation 
of the electron to proton mass ratio~\cite{ZelevinskyPRL08} and has already 
resulted in the observation of unusual non-adiabatic
effects~\cite{McGuyerPRL13,McGuyerNatPhys14}. Strontium molecules consisting of
even-isotope atoms, such as $^{88}$Sr$_2$ or $^{86}$Sr$^{88}$Sr, for
which the nuclear spin is zero, can only be formed by
photoassociation. Thus non-resonant light control of shape resonances
is particularly promising in this case~\cite{GonzalezPRA12}. The
amount of intensity that is required to achieve such control is
expected to depend on the field-free scattering length. 
The  scattering length is very small for $^{88}$Sr$_2$, and large for
$^{86}$Sr$^{88}$Sr, allowing a comparison of the intensity dependence
for the two limiting cases. All of these facts together make 
the strontium dimer a natural benchmark for our asymptotic model. 

The paper is organized as follows: 
We briefly recall the model for a diatomic molecule interacting with
non-resonant light in Sec.~\ref{sec:theory}. Introducing reduced
units of length and energy, we derive in Sec.~\ref{subsec:universaleq} 
a universal asymptotic Hamiltonian for this interaction. 
The nodal line technique to solve the corresponding asymptotic
Schr\"odinger equation is introduced in 
Sec.~\ref{subsec:nodallines:multi}, with the computational details
summarized in Appendix~\ref{app:nodal}. For the example of
$^{88}$Sr$_2$, we compare the results obtained from the asymptotic
model with the nodal technique to those obtained from diagonalization
of the full Hamiltonian (Sec.~\ref{subsec:88Sr}).  
The differences in field-dressed shape resonances for molecules with
small and large  scattering lengths are illustrated in
Sec.~\ref{subsec:86-88Sr}, for $^{88}$Sr$_2$ and $^{86}$Sr$^{88}$Sr
molecules. We conclude in Sec.~\ref{sec:conclusion}. 

\section{Interaction of a diatom with a non-resonant optical field}
\label{sec:theory}
The Hamiltonian of an atom pair in its electronic ground state in the
presence of a non-resonant laser field, assuming the Born-Oppenheimer
approximation,
is written in the molecule-fixed frame as 
\begin{equation}
  \label{eq:2D_Hamil}
  H =   T_R+\frac{{\mathbf{L}}^2}{2\mu  R^2}+V_g(R)
  -\frac{2\pi I}{c}\left(\Delta\alpha(R)\cos^2\theta+\alpha_\perp(R)\right)\,.
\end{equation}
In Eq.~\eqref{eq:2D_Hamil}, $T_R$ and $\mathbf{L}^2/2\mu R^2$ are the
vibrational and rotational kinetic energies for the motion of the two
nuclei with reduced mass 
$\mu$, interacting at interatomic separation $R$ through the
potential $V_g(R)$. The last term of Eq.~\eqref{eq:2D_Hamil},
where $c$ denotes the speed of light, represents the interaction with
non-resonant light of intensity $I$, linearly polarized along
the space-fixed $Z$ axis. $\theta$ denotes the polar angle between
the molecular axis and the laser polarization. The molecular
polarizability tensor is characterized by its  perpendicular and
parallel components $\alpha_\perp(R)$ and $\alpha_\parallel(R)$, 
determined with respect to the molecular axis,  which give rise to the 
polarizability anisotropy,
$\Delta\alpha(R)=\alpha_\parallel(R)-\alpha_\perp(R)$. 
  Note that the tensor $\alpha$, which has the dimension of a volume
  (cm$^3$ in cgs units), is related to the polarizability
  $\underline{\alpha}$ which is deduced from the induced  
  dipole moment (expressed in SI units  of CV$^{-1}$m$^{2}$) by
  $\underline{\alpha}=4\pi\epsilon_0\alpha$ with $\epsilon_0$ the
  vacuum polarizability. 
In Eq.~\eqref{eq:2D_Hamil}, the frequency of the 
non-resonant light is assumed to be far detuned from any resonance 
which allows for using the static polarizabilities.
A large effect of the non-resonant light is expected if the
light-matter interaction strength is large compared to the rotational
kinetic energy.  This corresponds to small rotational
constant, or large reduced mass, and to large atomic polarizabilities.

The long-range behavior of the $R$-dependent 
polarizability, valid at $R>R_d=(4\alpha_1\alpha_2)^1/6$, 
can be derived from the polarizabilities of the two constituent atoms,
$\alpha_1$ and $\alpha_2$. 
In the electronic ground state, one obtains~\cite{HeijmenMP96,JensenJCP02}
\begin{subequations}
  \label{eq:a_long}
  \begin{eqnarray}
    \alpha_\parallel(R)&\approx&\alpha_1+\alpha_2+\frac{4\alpha_1\alpha_2}{R^3}
    +\frac{4(\alpha_1+\alpha_2)\alpha_1\alpha_2}{R^6} \label{a_para_long} \,,\\ 
    \alpha_\perp(R)&\approx&\alpha_1+\alpha_2-\frac{2\alpha_1\alpha_2}{R^3}
    +\frac{(\alpha_1+\alpha_2)\alpha_1\alpha_2}{R^6} \label{a_per_long}   \,.
  \end{eqnarray}  
\end{subequations}
This $R$-dependence needs to be connected to \textit{ab initio} data
at short range. If this data is not available for the molecule of
interest, the parallel and perpendicular polarizability components can be
approximated (as in the present paper) by keeping them constant for $R<R_C$ 
and employing Eqs.~\eqref{eq:a_long} for $R>R_C>R_d$. The last inequality 
avoids the divergence occuring in $\alpha_\parallel$ at $R_d$.

The non-resonant field introduces a mixing of different partial waves of
the same parity such that $\ell$ is not a good quantum number. For a given 
diatom, the rovibrational 
levels and low-energy scattering states can be determined by solving
the Schr\"odinger equation associated to 
the Hamiltonian~\eqref{eq:2D_Hamil}. To this end, $H$ is represented by 
a mapped grid for the radial part~\cite{WillnerJCP04} and a basis set 
expansion in terms of Legendre polynomials $P_{\ell}(\cos\theta)$ 
for the angular part~\cite{abramowitz64}, taking advantage of the magnetic 
quantum number $m$ being conserved. We label the field-dressed states by the
field-free quantum numbers $\ell$, $m$ and $v$, adding a tilde to
indicate that they are labels not quantum numbers. For the bound
states, the field-dressed levels $\tilde v$, $\tilde\ell$ are
diabatically connected to the field-free quantum numbers even for
very high intensities. 

\section{Asymptotic model}
\label{sec:theory_asympt}
We derive an asymptotic approximation to the
Hamiltonian~\eqref{eq:2D_Hamil} by 
extending the  nodal line asymptotic model of Ref.~\cite{LondonoPRA10}
to account for the interaction of the diatom with a non-resonant field.
This is possible since the
influence of the non-resonant field on low temperature scattering
states and weakly bound levels is dominated by the long range part of
the interaction, characterized by a $1/R^3$-behavior
(see Eq.~\eqref{eq:a_long}) and since the resonances under study are
sufficiently 
close to the threshold. This method yields an efficient approach
to study near threshold properties, such as shape resonances, of a 
diatomic molecule subjected to an intense non-resonant field. 

\subsection{Universal asymptotic Schr\"odinger equation for a diatom
  interacting  with a non-resonant field}
\label{subsec:universaleq}
To derive the asymptotic approximation, we consider the Schr\"odinger
equation with $V_g(R)$ replaced by its leading order asymptotic term,
$V_g(R)\approx - C_6/R^6$ describing the van der Waals
interaction. For the interaction with the non-resonant field,
we also account only for the leading order term which scales as
$1/R^3$. In addition, the $R$-independent term in 
$\alpha_\perp(R)$, which reduces to $E_0=-\frac{4\pi }{c}\alpha_0I$, 
lowers the dissociation limit.  Taking
advantage of $m$ being conserved, 
the asymptotic 2D-Schr\"odinger equation reads
\begin{eqnarray}
  \label{eq:SE}
&&\left[-\frac{\hbar^2}{2\mu}\frac{d^2}{dR^2} - \frac{C_6}{R^6} 
+ \frac{\hbar^2}{2\mu}\frac{{\mathbf{L}}^2}{R^2}\right]
\psi(R,\theta)\\ \nonumber 
&&-\frac{2\pi I}{c}\left[2\alpha_0 + \frac{2\alpha_0^2}{R^3}(3\cos^2\theta-1) 
\right]\psi(R,\theta)
= E\psi(R,\theta)\,.  
\end{eqnarray}
Equation~\eqref{eq:SE} can be rescaled by introducing a dimensionless
reduced length $x$, a reduced energy $e$ (defined with respect to the
field shifted dissociation limit $E_0$) and a reduced  laser field
intensity $i$,
\begin{eqnarray*}
R & = & \sigma x\,, \\
E - E_0& = & \epsilon \,e\,, \\
I& = &\beta~i \,.
\end{eqnarray*}
The unit conversion factors for  length $\sigma$, energy $\epsilon$
and laser intensity  $\beta$ contain the information specific to the
free molecule:
\begin{subequations}
\label{eq:scaling}
\begin{eqnarray}
\sigma & = & \left(\frac{2\mu C_6}{\hbar^2}\right)^{1/4}\,, \\ 
\epsilon & = & \frac{\hbar^2}{2\mu\sigma^2}\,, \\ 
\beta & = & \frac{c}{12\pi} \frac{\hbar^{3/2}C_6^{1/4}}{\alpha_0^2(2\mu)^{3/4}} 
=\frac{c\sigma^3\epsilon}{12\pi\alpha_0^2}\,.
\label{eq:beta}
\end{eqnarray}
\end{subequations}
A unit conversion factor for time is obtained from that of energy, 
$\tau=\hbar/\epsilon$. 
The unit conversion factor for intensity,  $\beta$, is 
proportional to $\alpha_0^{-2}$ and to $\mu^{-3/4}$, such that larger
polarizability and larger reduced mass require less intensity $I$ for
achieving the same value of the reduced intensity $i$ (if the atoms
are not identical, $\alpha_0^2$ simply needs to be replaced by 
$\alpha_1\alpha_2$, see Eq.~\eqref{eq:a_long}). Similarly, 
since $\beta$ increases with $C_6^{1/4}$, atom pairs interacting
through weak van der Waals interaction are more sensitive to laser
field effects than those with strong interaction. 

Employing atomic units, that is Bohr radii $a_0$ for $\sigma$,
Hartree for $\epsilon$ and $a_0^3$ for the atomic polarizability,  and
expressing the laser intensity $I$ in GW/cm$^2$, the reduced intensity
is given by
\begin{equation}\label{eq:scale:I}
  i=4.274177 ~ 10^{-8} ~ \frac{\alpha_0^2 I}{\epsilon \sigma^3}\, ,
\end{equation}
whereas the shift of the dissociation limit in reduced units is equal to
\begin{eqnarray}\label{eq:scale:E0}
e_0=\frac{E_0}{\epsilon}=-\frac{4\pi \alpha_0 I}{c \epsilon}
=-1.424725 ~ 10^{-8} ~ \frac{\alpha_0 I }{\epsilon}\,. 
\end{eqnarray}
When the reduced energy is expressed in $\mu$K, the numerical factors
are equal  to 13496.717 for the reduced intensity and 
-4498.93 for the reduced threshold shift, respectively. 
The asymptotic 2D-Schr\"odinger equation in reduced units is given by
\begin{equation}
  \label{eq:asy}
  \left[-\frac{d^2}{dx^2} - \frac{1}{x^6} + \frac{\mathbf{L}^2}{x^2}
    - i
    \frac{\cos^2\theta -1/3}{x^3}
    - e
  \right] f (x,\theta) = 0\,.
\end{equation}
The asymptotic Schr\"odinger equation is valid at sufficiently large
distances where the potential is dominated by the $1/x^6$ term,
i.e.,  for $x>x_{asym}=(C_8/C_6)^{1/2}$.  
To solve the asymptotic Sch\"odinger equation for $x>x_{asym}$,
we introduce below 
a modification of the nodal line technique which accomplishes this task.

The asymptotic model in reduced units predicts that a field-free 
shape resonance is solely determined by Eq.~\eqref{eq:asy}, i.e. by its 
rotational quantum number $\ell$, and by boundary 
conditions at short distance, $x_{0\ell}>x_{asym}$, which are related 
to the value in reduced units of
the $s$-wave scattering length of the molecule. In the
presence of a non-resonant field, the resonance energy (in reduced
units) depends, apart from the field-free scattering length, on both the reduced
laser field intensity $i$ and the field-free rotational quantum number $\ell$.

\subsection{Nodal line technique}
\label{subsec:nodallines:multi}
In order to solve Eq.~\eqref{eq:asy}, we expand 
the wave function in Legendre polynomials, $P_{\ell}$
\begin{equation}
  \label{eq:angmom}
  f(x,\theta) = \sum_\ell y_{\ell}(x)
  P_{\ell}(\cos\theta)\,, 
\end{equation}
introducing the radial functions $y_{\ell}(x)$ for the
different coupled channels $\ell$. 
Eq.~(\ref{eq:asy}) is then replaced by a system of coupled equations
which can be written in vectorial form,
\begin{equation}
  \label{eq:asyvect}
 \frac{d^2}{dx^2}{\bf y}(x) + ({\bf M}+ e~{\openone}) {\bf y}(x)= 0\,,
\end{equation}
where the vector ${\bf y}(x)$ is the set of functions $y_{\ell}(x)$,
$\openone$ denotes the identity and $\bf{M}$ is the matrix of the
operator $\frac{1}{x^6} - \frac{{L}^2}{x^2} +
i\frac{\cos^2\theta-1/3}{x^3}$ represented in the basis of 
Legendre polynomials with $\ell$-values of the same parity. 
We restrict here to $m=0$ and even $\ell$ values varying from 0 to
various $\ell_{max}$, i.e., the model consists of $n=\ell_{max}/2+1$
channels $\ell=0,\,2,\,...,\,2(n-1)$. We denote by ${\bf y}^j(x)$ a
particular solution of the asymptotic Schr\"odinger equation in the
coupled channel model, 
\begin{equation}
  \label{eq:decomp}
  {\bf y}^j(x) = \sum_{\ell\,\mathrm{even}\,=0}^{\ell_{max}}
  y_{\ell}^j(x) |\ell, 0\rangle\,, 
\end{equation}
where $y_{\ell}^j(x)$ is the radial component of the $j$th solution
in the $\ell$th channel.

With the nodal line technique, the solution of the coupled
equations~\eqref{eq:asyvect} is performed only in the asymptotic zone, 
where the asymptotic Hamiltonian is valid. At large distance, physical
boundary conditions are imposed, depending on the sign of $e$. For
$e<0$, the radial wave functions exponentially decay in all channels,
quantifying  the energy of  bound levels. For $e\ge0$, regular and
irregular Bessel functions characterize the asymptotic behavior. At
small distance, on the frontier of the inner zone, we require the
radial part of the physical wave function in each channel $y_{\ell}^j$  
to vanish at a position that is located on a $\ell$-dependent straight 
line in the $(e,x)$ plane, the so-called nodal
line~\cite{VanhaeckePhD,CrubellierEPJD99}.  
Without non-resonant field, the following positions were
used~\cite{VanhaeckePhD,CrubellierEPJD99}:
\begin{equation}
  \label{eq:nodalline}
  {x}_{0\ell}=x_{00}+Ae+B\ell(\ell+1)\,,
\end{equation}
where the parameters $x_{00}$, $A$ and $B$ are characteristic of the chosen
atom pair. In particular, $x_{00}$ corresponds to the position of a
node of the 
threshold $s$-wave wave function and is related to the $s$-wave
scattering length~\cite{CrubellierJPB06}. $A$ takes  the
variation  of the node position with energy in the wave function with 
$\ell=0$ into account. $B$ describes the shift in the node of the
threshold wave functions induced by the centrifugal term for the
various partial waves, $\ell>0$. The parameters $x_{00}$, $A$ and $B$
are adjusted, if possible, to experimental data, such as the positions
of bound levels or resonances  close to threshold, and the $s$-wave
scattering length. They can also be calculated from molecular
potentials, when available.   

In the absence of either reliable potentials or experimental data,
there is a rough, but universal estimate of these parameters given by
very simple analytical formulas which depend only on the $s$-wave
scattering length, $A^G=-(x_{00})^7/8$ and $B^G=(x_{00})^5/4$
\cite{LondonoPRA10}. These laws are deduced from the universal model
of Ref.~\cite{GaoPRA98} which consists in a  $-1/x^{6}$
potential limited by an infinite repulsive wall at a distance
$x_{0_{G}} \rightarrow 0$. The WKB approximation it used to evaluate,
in the vicinity of the threshold  and for a not too high value of
$\ell$, the shift of the node  located at $x_{00}$ that arises from
the contribution of the kinetic $Ae$ and centrifugal ${B\ell(\ell+1)}$
energies in the range $x_{0_{G}}\le x \le x_{00}$
\cite{VanhaeckePhD}. Although  the 
model becomes less realistic as $x_{00}$ decreases, the corresponding
$A^G$ and $B^G$ values are comparable to the values ajusted to
experimental data \cite{CrubellierEPJD99, PasquiouPRA10}.   

In the presence of a laser field, as it will be shown below (see 
Sec.~\ref{subsec:88Sr}), an intensity dependent term has to be added
to the nodal lines, 
\begin{equation}
\label{eq:nodalline_i}
{x}_{0\ell}=x_{00}+Ae+B\ell(\ell+1)+Ci\,.
\end{equation}
The new term, $Ci$, i.e., lowest order in $i$, accounts for the
contribution of the interaction with the non-resonant field at short
range.  With this modification it is possible to obtain the
the bound levels, the resonance profiles of the shape resonances as
well as the scattering length of the field dressed molecule for any
intensity. We mention here that the {\it i}-dependent 
term can be evaluated in the same way as $A^G$ and $B^G$. It is 
even possible to use exactly the same description of the polarizability 
as in the full-potential calculations (see section
\ref{sec:strontium}): using the 
diagonal term of Eq.~(\ref{eq:asy}) (in which we have replaced $\cos^2\theta$
by its approximate $\ell$-independent mean value 1/2) 
for $x>x_C=R_C/\sigma$ and kept a constant polarizability for $x \le
x_C$), 
we obtain (in reduced units)  
\begin{equation}
\label{eq:C_G}
C^G=-x_{00}^4/12+3 x_{C}^4/48.
\end{equation}
In order to determine bound levels and resonances, 
Eq.~\eqref{eq:asyvect} is prealably solved numerically by inward
integration starting from a large value $x_\infty$, imposing only 
large $x$ boundary conditions. For $e<0$, i.e.,
levels below threshold, this
value has simply to be larger than the outer Condon point. For $e\ge0$,  
$x_\infty$ is chosen in the $x$-domain where the diagonal elements of 
the matrix ${\bf M} +e \openone$ reach their asymptotic form, that is 
$[e-\ell(\ell+1)/x^{2}]$ for $e > 0$ and $[1/x^6-\ell(\ell+1)/x^{2}]$
for $e=0$. One can then use analytical solutions, i.e., Bessel
functions, as initial values for the inward integration of the radial
functions $y_\ell^j(x)$ in each channel~\cite{MoritzPRA01} and
construct a set of linearly independent solutions ${\bf y}^j$ with the
correct asymptotic  behavior. There are $n$ such solutions for bound
levels and Siegert states. Their asymptotic behavior corresponds to 
either an exponentially decreasing function or an outgoing complex wave
function in a given channel and zero in all others. For 
$e\ge0$, there are $2n$ linearly independent solutions, with an
asymptotic behavior given by either a regular or an irregular Bessel
function in a given channel and zero in all others. 

The physical solutions ${\bf z}^k$ are linear combinations of the
particular solutions ${\bf y}^j$ prealably calculated. 
The coefficients are determined by
imposing the radial components in each channel to vanish at the
corresponding node position $x_{0\ell}$. This short range condition
leads to a quantization of energy for the bound levels and Siegert
states. It also allows to determine the scattering length in the
presence of the non-resonant field~\cite{AM_SL}. 

The continuous, $n$-fold degenerate spectrum at an energy $e$
is described by Multichannel Scattering Theory
\cite{seaton83,fano86,smithPR60}.  
The chosen asymptotic boundary conditions allow for a direct determination
 of the energy-dependent reaction matrix, ${\bf K}(e)$, 
from which the scattering matrix, ${\bf S}(e)$, and the time-delay
matrix, ${\bf Q}(e)$, are easily deduced. The details are presented in
Appendix~\ref{app:nodal}. In particular, the  $\bf Q$ matrix is
well-adapted to analyze shape resonances, by studying the
energy variation of its lowest eigenvalue $q_1(e)$ which corresponds
to Lorentzian profiles, see Appendix~\ref{app:subsec:profiles} for details.
The eigenvalues $\tan[\tau_j]$ of the $\bf K$ matrix allow for determining 
the eigenphase sum $\tau(e)$. The energy variation of the derivative 
of the eigenphase sum yields also a profile of the shape resonances. 
The resonances are also finally characterized by calculating the
energy variation of either population or mean value of  $1/x^2$ 
inside the rotational barrier, see Appendix~\ref{app:subsec:profiles} 
and~\ref{app:subsec:siegert} for details. 

\section{Shape resonances in strontium}
\label{sec:strontium}

We investigate here the shape resonances of two isotopomers of 
strontium, $^{88}$Sr$_2$ and $^{86}$Sr$^{88}$Sr. They have the largest
natural  abundances (68\% and 16\%) and no nuclear spin.
The $s$-wave scattering lengths are $a_S=-2\,a_0$, or -0.013 in reduced 
units, for $^{88}$Sr$_2$ \cite{SteinEPJD10} and  $a_S=97.9\,$a$_0$, or
0.664 in reduced units, for $^{86}$Sr$^{88}$Sr \cite{ZhangCPL11} 
(see Table \ref{tab:redunits:88Sr} for the scaling factors).
For close to zero scattering length, quantum defect theory predicts 
shape resonances with $\ell=4, 8, 12,\ldots$ (i.e., for the case of
$^{88}$Sr$_2$), whereas for a large
scattering length, i.e., for $^{86}$Sr$^{88}$Sr, 
shape resonances with $\ell=2, 6, 8,\ldots$ are
expected~\cite{GaoPRA09}. We first test the validity of the asymptotic
model by comparing to exact results for $^{88}$Sr$_2$ and then compare
the behavior of the shape resonances as a function of the non-resonant
light intensity for the two isotopomers.
\begin{table}[tb]
  \centering
  \begin{tabular}{|l|c|c|c|c|}
    \hline 
    &$\sigma\,$[a$_0$]& $\epsilon\,[\mu$K$]$ & 
    $\beta\,[$GW/cm$^{2}]$ & $\tau\,$[ns]\\\hline
   $^{88}$Sr$_2$ &
    151.053 & 
    86.3653 & 
    0.635782 &
    88.4409 \\ \hline
    $^{86}$Sr$^{88}$Sr & 
    150.617 & 
    87.876 & 
    0.641319 &
    86.9204 \\ \hline
  \end{tabular}
  \caption{ Scaling factors defining the reduced units,
    cf. Eq.~\eqref{eq:scaling}, for 
    $^{88}$Sr$_2$ and $^{86}$Sr$^{88}$Sr, obtained
    for $C_6=3246.97\,$a.u. and $\alpha_0=186.25\,$a$_0^3$.
  } 
  \label{tab:redunits:88Sr}
\end{table}

\subsection{Validity  of the asymptotic model: Position, width and
  hybridization of shape resonances in $^{88}$Sr$_2$} 
\label{subsec:88Sr}

To test the validity  of the asymptotic model, we solve the asymptotic
Schr\"odinger equation~\eqref{eq:asy} and compare to results 
of the full 2D Hamiltonian of Eq.~\eqref{eq:2D_Hamil}, using 
the ground state potential energy curve from
Ref.~\cite{SteinPRA08}, adjusted to yield the relevant scattering length.
The polarizabilities are computed from Eq.~\eqref{eq:a_long} for
$R>R_C=10$~a$_0$ with an atomic polarizability of
$\alpha_0=186.25\,$a$_0^3$~\cite{CRChandbook}; 
for $R\le R_C$ the polarizabilities are taken to be constant.
We first need to determine the nodal lines.  
To this end we use Ref.~\cite{LondonoPRA10} which gives
the energies and widths of shape 
resonances as a function of the position of a node at short
range. Reversely, knowing the position of a field-free shape
resonance, it is possible to find a node position $x_{0\ell}$ (in a
chosen $x$-interval) that yields a resonance at this energy value.  
Starting from the field-free positions of the shape
resonances $\ell$=4, 8, 12 and 16~\cite{GonzalezPRA12}, we first test  
nodal lines of the type~\eqref{eq:nodalline}. 
Since the coefficient $A$ in Eq.~\eqref{eq:nodalline}
plays a minor role,  it is taken to be constant and equal to
$A=A^G=-(x_{00})^7/8$, the value of the 'universal' model 
\cite{LondonoPRA10}. $B(\ell)$ is taken to be a polynomial of degree 3 
in $\ell(\ell+1)$; $x_{00}$ and $B(\ell)$ are determined by a fit to
the field-free shape resonance positions
(the degree of the polynomial is 3 to fit the 4 data points exactly). 
Note that this fit provides the correct value, 
$a=-2a_0$, of the field-free scattering length. 

However, when using the ansatz~\eqref{eq:nodalline} to determine, in addition 
to the field-free positions of the shape resonances, the slopes of their
dependence on the non-resonant field, the result is disappointing: For 
the four resonances the slopes are smaller by a factor of 
approximately 1.75 compared to  those obtained from the full
Hamiltonian.
This finding suggests that the contribution of the short-range 
part of the interaction with the non-resonant field (for
$x<x_{0\ell}$) is
non-negligible, rendering the use of field-independent nodal lines 
insufficient. Remarkably, the effect of the coupling at 
short range on the intensity dependence of the
resonance positions can be simply compensated, at least roughly, by
introducing a scaling factor in the field intensity. 

\begin{figure}[tb]
  \centering
  \includegraphics[width=0.85\linewidth]{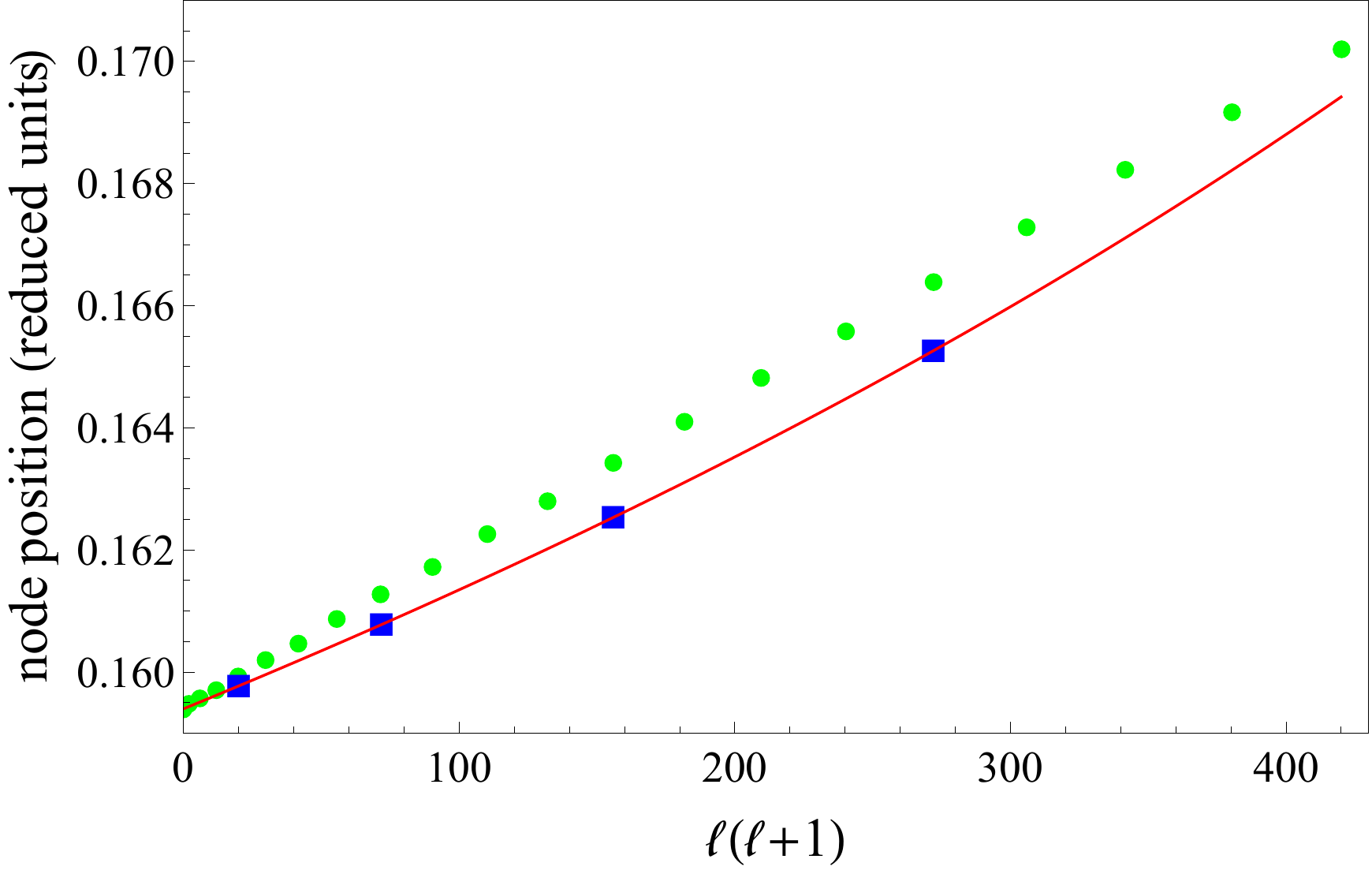}
  \caption{\label{fig:noeuds-zi} Nodal lines 
    for $^{88}$Sr$_2$ 
    in reduced units: The red line shows the variation  of the node
    positions at threshold with $\ell(\ell+1)$ for $i=0$, compared to
    the 'universal' 
    model corresponding to the same value of $x_{00}$ (green dots). 
    The blue squares indicate the values for 
    $\ell=4,8,12,16$, adjusted to reproduce the field-free positions
    of the corresponding shape resonances~\cite{GonzalezPRA12}). 
    The variation from $i=0$ to $i=1$ (in reduced units) of the nodes 
    is too small to be visible in this figure. 
  } 
\end{figure}

The influence of the interaction with the
non-resonant field at short range 
on the resonance positions can be fully accounted for in
the asymptotic model by making the nodal lines intensity-dependent,
cf. Eq.~\eqref{eq:nodalline_i}. Assuming $A$ to be constant, $A=A^G$,
as above,  $x_{00}$ and  $B(\ell)$, taken to be a  polynomial of
degree 4 in $\ell(\ell+1)$, are adjusted to reproduce exactly
the nodes of the field-free wave functions with 
$\ell=0,4,8,12,16$. As above, this fit provides the correct value, 
$a=-2a_0$, of the field-free scattering length. Additionally, 
$C$, taken to be a polynomial of degree 3 in $\ell(\ell+1)$, is
adjusted to reproduce exactly the variation of the node positions for  
$\ell =4,~8,~12,~16$ with intensity when $i$ is increased from 0 to
1~reduced 
unit. To this end, the $\ell$-wave function of the rovibrational level
closest to threshold is obtained numerically 
for $i=0$ and $i=1$, employing the Fourier grid method to solve 
Eq.~(\ref{eq:2D_Hamil}) in a single channel approximation. 
For $i=1$, the single channel calculation represents an approximation.
It is, however, well justified by the very small $\ell$ mixing 
observed in a coupled channels calculation for $i=1$, corresponding to
$I=0.64\,$GW/cm$^2$ for $^{88}$Sr$_2$. 
The variation of the node positions at threshold with $\ell(\ell+1)$ 
is shown in Fig.~\ref{fig:noeuds-zi} for $i=0$. Also plotted
are the node positions at threshold of 
the 'universal' model, i.e., $B(\ell)=B^G=x_{00}^5/4$, corresponding
to the same value of $x_{00}$. They do not deviate much from the node
positions obtained from the full potential, except for large $\ell$
values.  The node positions in the presence of a weak non-resonant
field, $i=1$ in reduced units,  differ from those for $i=0$ by only 
about $-0.5 \times 10^{-4}$~reduced units, not visible on the scale of
the figure.
\begin{figure}[tb]
  \centering
  \includegraphics[width=.99\linewidth]{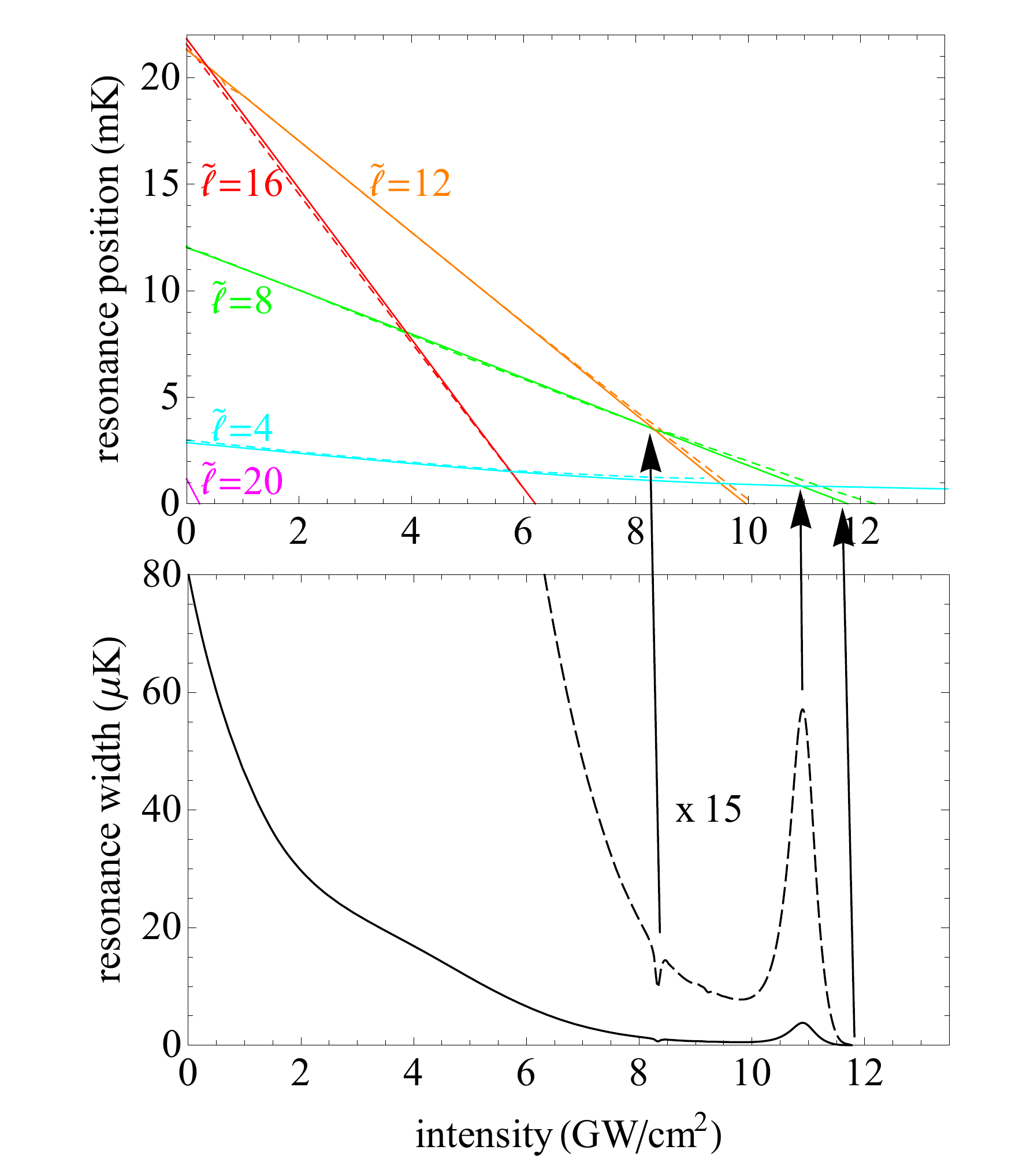}
  \caption{\label{fig:comparison} Upper panel: Position of the shape
    resonances of $^{88}$Sr$_2$ as a function of non-resonant field intensity
    ($\tilde \ell=4$ (cyan), $\tilde \ell$=8 (green), 
    $\tilde \ell =12$ (orange), $\tilde \ell$=16 (red) and $\tilde
    \ell$=20 (magenta)), obtained from  the asymptotic
    model with intensity-dependent nodal lines (solid lines) and the
    full Hamiltonian (dotted lines) -- the results 
    are almost indistinguishable. 
    Lower panel: Width of the  $\tilde \ell=8$ shape resonance of
    $^{88}$Sr$_2$ as a function of non-resonant field intensity.
    The dashed curve displays a 15-fold zoom 
    of the solid one, showing the broadening (resp. narrowing) of the 
    ${\tilde \ell}=8$ resonance when it crosses the ${\tilde \ell}=4$ 
    (resp. ${\tilde \ell}=12$) one, as indicated by the arrows. At
    threshold, also indicated by an
    arrow, the width tends to zero as expected. 
  } 
\end{figure}
Adding this small and simple linear intensity dependence to the nodal lines
yields spectacular agreement of the asymptotic model with the full
Hamiltonian. This is demonstrated by the upper panel of
Fig.~\ref{fig:comparison} which compares the results of the asymptotic
model with intensity-dependent nodal lines to those of the full
Hamiltonian: Almost no difference is visible on the scale of the 
figure. A linear intensity dependence of the nodal lines thus allows
for utilizing the asymptotic model up to very large field
intensities. 

Note that all crossings between resonances or levels in
Fig.~\ref{fig:comparison} are in fact avoided crossings, and the 
diabatized lines are simply labeled by the 
${\tilde \ell}$ value equal to the field-free $\ell$ value. 
Figure \ref{fig:comparison} also shows the behavior of the resonance
width as a function of field intensity for the example of 
${\tilde \ell}=8$ (lower panel). 
The calculations using the asymptotic nodal line technique were
performed  with 11 coupled channels, but we have checked for several 
values of ${\tilde \ell}$ and $i$ that the positions of the shape resonances (up
to ${\tilde \ell}=20$) do not change when $\ell _{max}$ is increased (up to
$\ell_{max}=24$, corresponding to 13 coupled channels). The resonance
positions and widths for ${\tilde \ell}=8,12,16$ shown in
Fig.~\ref{fig:comparison} have been obtained with the complex energy
method, cf. Appendix~\ref{app:subsec:siegert}.  For the resonance
${\tilde \ell}$=4, which is close to the top of the corresponding  
barrier at $i=0$, the complex energy method does not apply and
resonance profiles  have been determined from the smallest eigenvalue
of the time-delay matrix $Q(e)$, cf. Appendix~\ref{app:subsec:profiles}. 

\begin{figure}[tb]
  \includegraphics[width=0.99\linewidth]{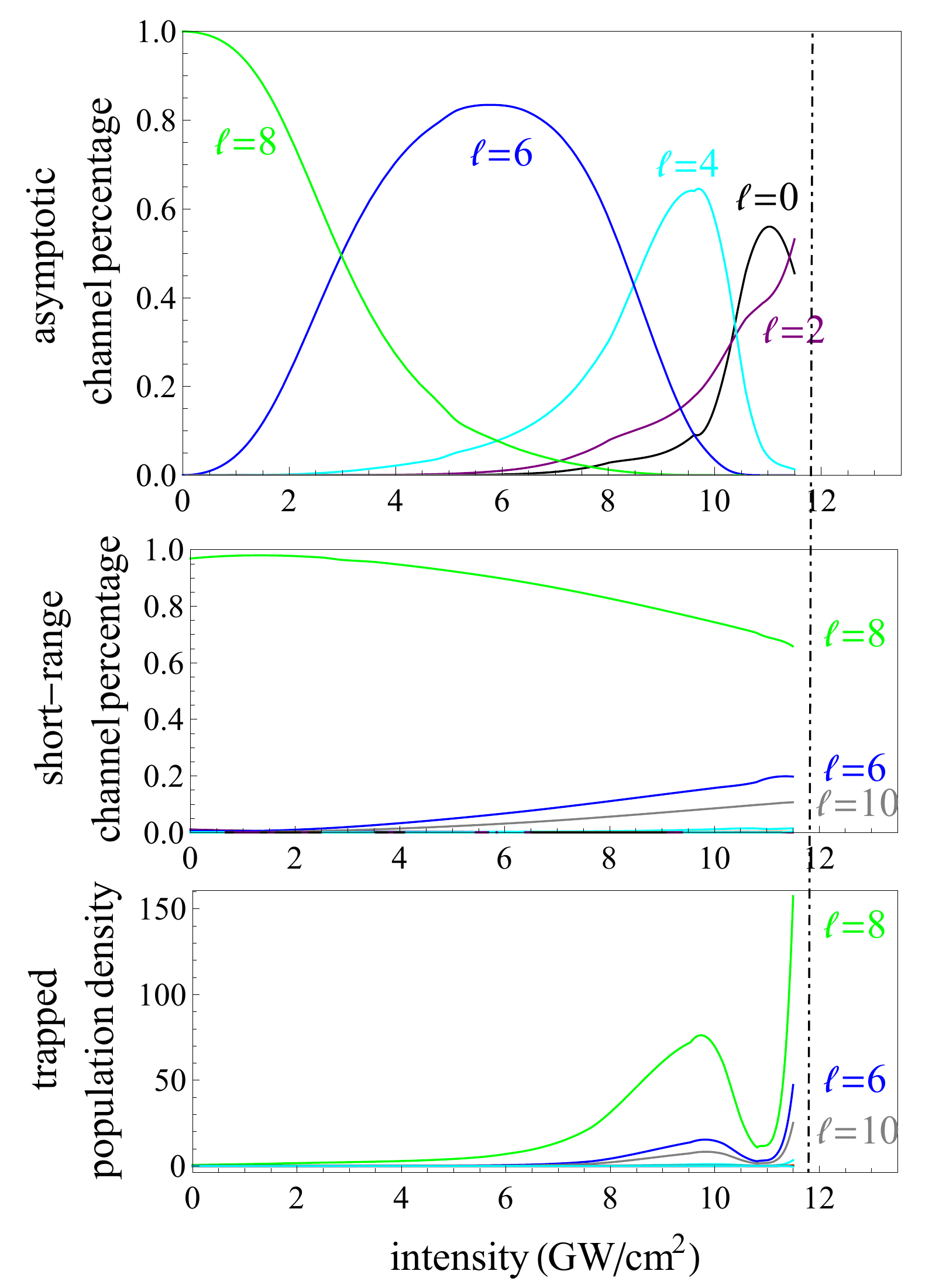}
  \caption{\label{fig:mixing-cont-zi} 
    $^{88}$Sr$_2$: Hybridization of the ${\tilde \ell}=8$ shape
    resonance. 
    The top panel shows the 'asymptotic' percentages in the different
    channels, i.e., the square of the partial wave components of the
    continuum wave function  
    associated to the lowest eigenvalue of the time delay-matrix. 
    The bottom panel displays 
    the population densities (per energy unit) in the different channels which 
    are trapped inside the corresponding rotational barriers; and the middle 
    panel shows the 'short-range' channel percentages obtained from the 
    population densities presented in the bottom panel. The dot-dashed line 
    indicates the crossing with threshold.} 
\end{figure}

The intensity dependence of the resonance positions and widths shown
in Fig.~\ref{fig:comparison} is related to a strong hybridization of
the rovibrational 
motion~\cite{Aganoglu11,GonzalezPRA12}. The hybridization involves 
different aspects, which can be analyzed from profile 
calculations, cf. Appendix~\ref{app:subsec:profiles}.
This is shown in Fig.~\ref{fig:mixing-cont-zi}, illustrating 
$\ell$-mixing for the example of the ${\tilde \ell}=8$ resonance. 
The population density (per energy unit) trapped behind the centrifugal 
barrier (lower panel of Fig.~\ref{fig:mixing-cont-zi}) 
is essentially always concentrated in  the $\ell =8$ channel. It
increases rapidly when approaching the threshold.  
The crossing with the ${\tilde \ell}=12$ resonance does not visibly
affect this evolution, whereas the crossing with ${\tilde \ell}=4$
involves a clear  
decrease of the population density in the $\ell =8$ channel
(dip in the green line near 11$\,$GW/cm$^2$).
Note the different behavior of the percentages at short and long
range. In the short-range region the population 
is essentially concentrated in the $\ell =8$ channel; and the 
short-range percentages converge exactly to the population percentages
in  the different channels of the corresponding ${\tilde \ell}=8$ bound
level at threshold (middle panel of Fig.~\ref{fig:mixing-cont-zi}). In
contrast, the asymptotic percentages (upper panel of
Fig.~\ref{fig:mixing-cont-zi}) are very different from the short-range 
ones, with a very small contribution of the
$\ell=8$ partial wave and large contributions of partial waves with
$\ell=0$ and 2 at high intensities. The asymptotic percentages
represent the partial wave decomposition of  
the continuum wave function associated to the lowest eigenvalue of the
time delay  
matrix $Q(E)$. We stress here that this wave function is, inside the multiply 
degenerate continuum, the only wave function exhibiting  resonant behavior. 
The behavior of the asymptotic percentages
would probably be important if dynamical 
processes were considered.

\subsection{Comparing $^{88}$Sr$_2$ and $^{86}$Sr$^{88}$Sr: 
  Intensity dependence of shape resonances in molecules 
  with small and large  scattering length} 
\label{subsec:86-88Sr}
The crucial free parameter in the asymptotic model, and the only free
parameter  in the 'universal' asymptotic model, is the value of the 
$s$-wave scattering length (in reduced units) which determines the
node positions. It is thus  
particularly instructive to compare the $^{88}$Sr$_2$ and $^{86}$Sr$^{88}$Sr 
dimers.
Since for small $\ell$ the differences between the 'universal' and the
'realistic' nodal lines are small, see Fig.~\ref{fig:noeuds-zi}, 
we use here 'universal' nodal lines~\eqref{eq:nodalline_i}
for $^{86}$Sr$^{88}$Sr, with
coefficients $A=A^G$, $B=B^G$, $C$=0 and the value of $x_{00}$ determined
by the $s$-wave scattering length. In this essentially
explorative work, we have also limited the  
number of channels to 5 ($\ell_{max}$=8), sufficient to study the
$\ell=2$ and $\ell=6$ resonances.

\begin{figure}[tb]
  \centering
  \includegraphics[width=0.95\linewidth]{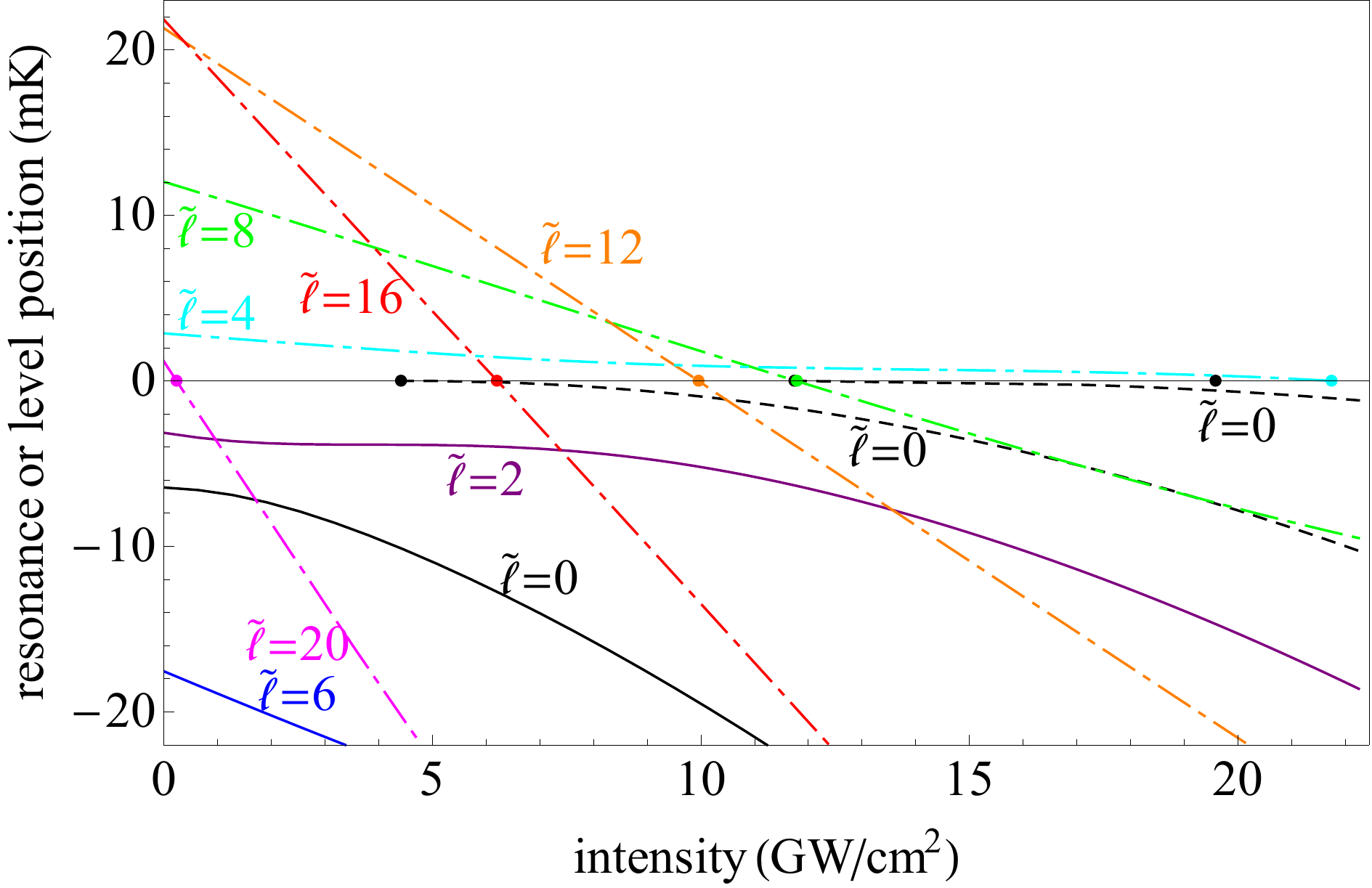}
  \caption{\label{fig:niveaux-zi} Positions of bound levels and shape
    resonances as a function of non-resonant field intensity for 
    $^{88}$Sr$_2$, calculated with realistic, $i$-dependent, nodal 
    lines (see text). Bound levels that are bound also in the field-free
    case are drawn as solid lines, dot-dashed lines correspond to 
    shape resonances which become bound at a certain intensity, 
    and 'supplementary' levels, i.e., regular scattering states that
    become bound, are represented by dashed lines. The colors correspond 
    to a 'diabatic' labeling. 
  }
\end{figure}

\begin{figure}[tb]
  \centering
  \includegraphics[width=0.95\linewidth]{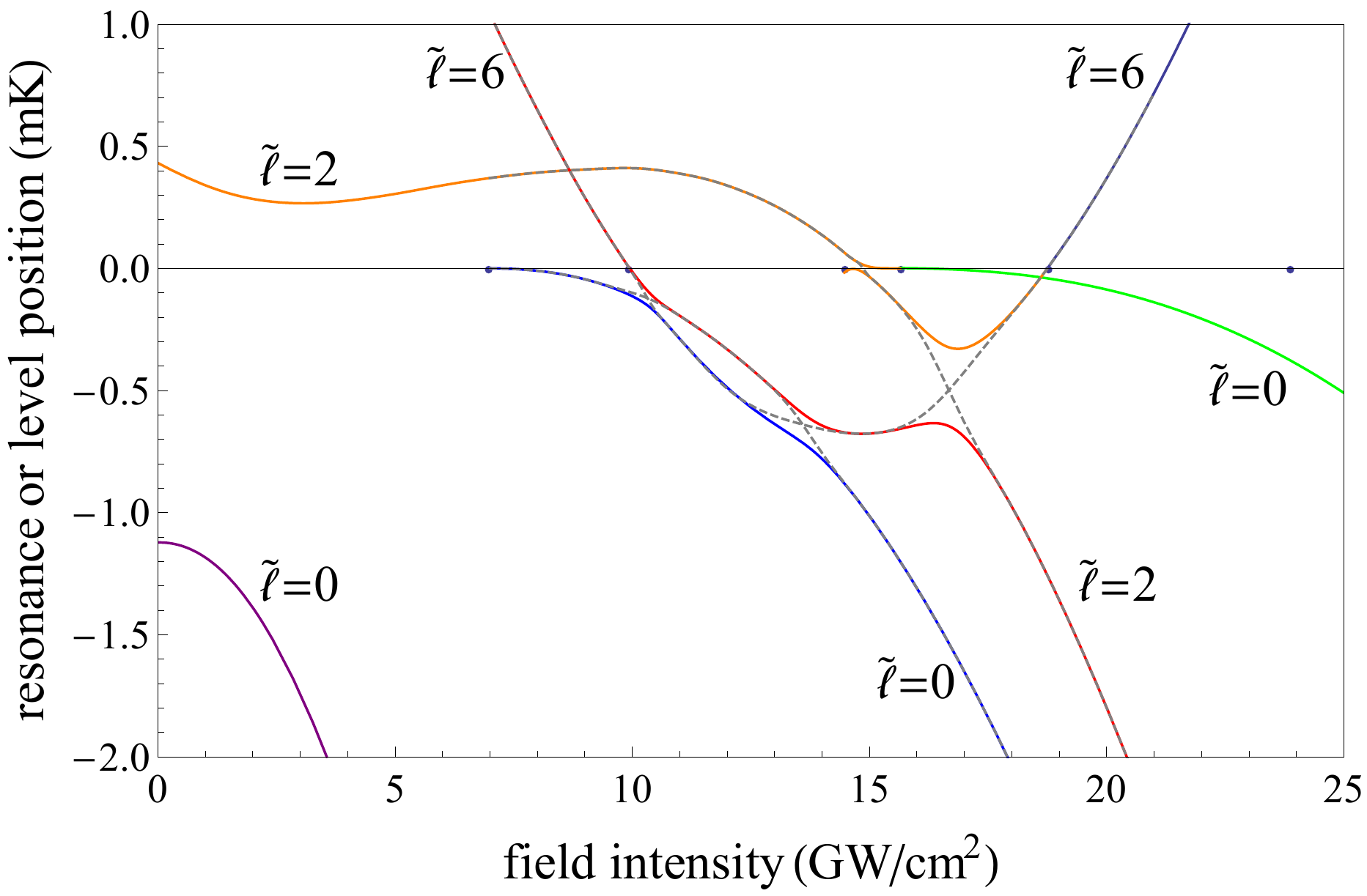}
  \caption{\label{fig:niveaux-8688} Positions of bound levels and shape
    resonances as a function of non-resonant field intensity for 
    $^{86}$Sr$^{88}$Sr, calculated with 'universal' 
    intensity-independent
    nodal lines. Avoided crossings are here clearly visible, making it 
    impossible to match colors and labels. Diabatized curves, 
    to which  the labeling corresponds, are drawn in black dashed lines.}
\end{figure}

Encouraged by the very good agreement between the asymptotic model and
the full Hamiltonian for the shape resonances, we calculate for both
isotopomers, in addition to the shape resonances, 
bound levels very close to threshold. 
Figure~\ref{fig:niveaux-zi} displays the positions of shape resonances 
and bound levels for $^{88}$Sr$_2$, 
whereas the corresponding results for $^{86}$Sr$^{88}$Sr 
are shown in Fig.~\ref{fig:niveaux-8688}. 
In both figures, we characterize each level or resonance by a value
${\widetilde \ell}$, with the labeling done by continuity through
avoided crossings ('diabatic' labeling) in two concording ways. First,
we observe how the levels appear, as the number of channels in the
calculations is enlarged; 
second, we analyze the channel decomposition for $i=0$. 
The intensity dependence of the resonance and bound level positions 
is extremely different for $^{88}$Sr$_2$ and     $^{86}$Sr$^{88}$Sr: 
In Fig.~\ref{fig:niveaux-8688}, new ${\tilde \ell}$ values, 
${\tilde \ell}=2$ and ${\tilde \ell}=6$, appear, all crossings are
widely avoided and the ${\tilde \ell =6}$  
resonance crosses twice the threshold. It is worth mentioning that the
theoretical  
energies obtained for the field-free bound $^{88}$Sr$_2$ levels with $\ell=0$ and
$\ell=2$  (-74.64~reduced units or -134.4$\,$MHz and -36.37~reduced 
units or  -65.5$\,$MHz) are in good agreement with the experimental
values of -136.7~MHz and -66.6~MHz for the $v=62$, $\ell=0$ and the
$v=62$, $\ell=2$ shape resonances, respectively~\cite{EscobarPRA08}. 
In the presence of the non-resonant field, the bound levels in
Fig.~\ref{fig:niveaux-zi}  
and Fig.~\ref{fig:niveaux-8688} are of three types: (i) 'pure' bound levels which are
bound also in the field-free case; (ii) bound levels which appear when a
shape resonance is pushed below threshold as the non-resonant field
intensity is increased; and (iii) 'supplementary' bound
levels, which start tangentially to the threshold, i.e., regular
scattering states that become bound as the field intensity is
increased. The latter are due to a deepening of the $\ell=0$ adiabatic
potential as $i$ increases. This effect is also observed for a very
strong static electric field coupling to a permanent dipole
moment~\cite{RosarioNJP09}. 

\begin{figure}[tb]
  \centering
  \includegraphics[width=0.99\linewidth]{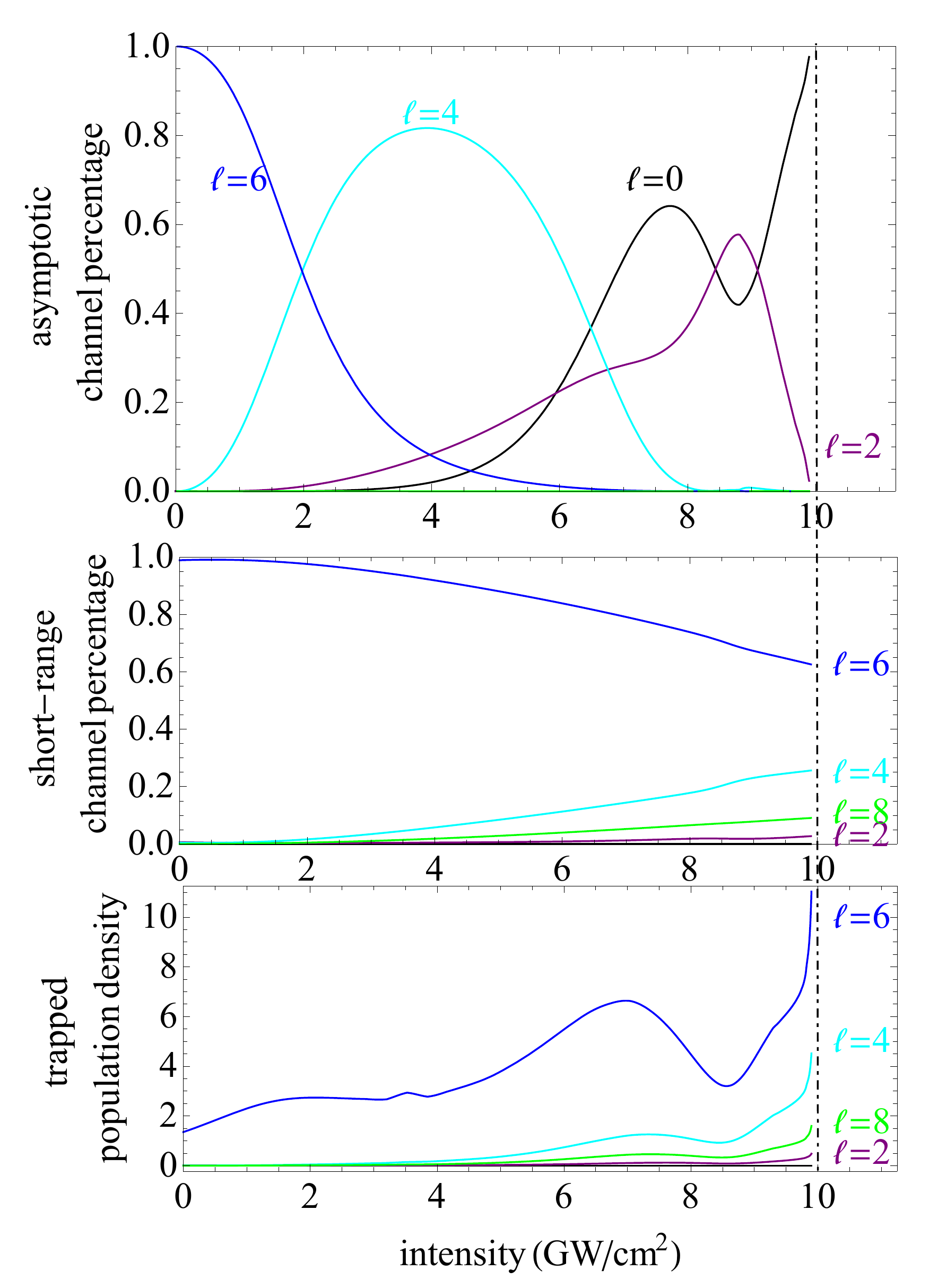}
  \caption{Hybridization of the ${\tilde \ell =6}$ shape resonance of
    $^{86}$Sr$^{88}$Sr as a function of non-resonant field intensity,
    analogously to  Fig.~\ref{fig:mixing-cont-zi}. 
    The dot-dashed line indicates the crossing with threshold.} 
  \label{fig:compare8688}
\end{figure}
Figure~\ref{fig:compare8688} analyzes the hybridization of a shape
resonance  for the example of the ${\tilde \ell=6}$ resonance in
$^{86}$Sr$^{88}$Sr. The features are similar  
to those shown in Fig.~\ref{fig:mixing-cont-zi}: In particular 
a drastic increase of the population density, especially in the 
$\ell=6$ channel, is observed when the resonance comes close to the threshold.
Simultaneously, there is almost no more contribution of the $\ell=6$ partial
wave in the asymptotic behavior.

\section{Conclusions}
\label{sec:conclusion}
We have generalized an asymptotic theory of diatomic scattering and
weakly bound molecular levels~\cite{GaoPRA09,LondonoPRA10} 
to account for the interaction of the
diatomic with non-resonant light through its polarizability
anisotropy. Solving the asymptotic Schr\"odinger equation by a nodal
line technique has allowed us to accurately reproduce the results of
the full Hamiltonian
for $^{88}$Sr$_2$ at all intensities.  
The asymptotic model thus allows for predicting the intensity
dependence of the positions and widths of shape resonances.

The field-free scattering length is the essential
parameter that determines the field-free position of shape resonances
and also the position of the
nodal lines. We have found an intensity dependence of the nodal line to be
required to accurately account for the effect of the polarizability
interaction at short range. Otherwise the slopes of the positions'
intensity dependence in the asymptotic
model differ by a factor of about 1.75 from those of the full
Hamiltonian. 
A similar factor appears in a single channel approximation to the
asymptotic model when intensity-independent nodal lines are
considered~\cite{AM_SingleCh}. 
The node positions are assumed to depend separately on energy,
rotational quantum number and non-resonant field intensity. 
The channel mixing is thus completely ignored at short range. 

The variation with field intensity of the resonance postions is found 
to be linear up to rather large field intensity.
This suggests the use of perturbation theory based on 
field-free properties only, i.e., a single-channel model.
A detailed discussion of such an approach will be presented in 
Ref.~\cite{AM_SingleCh}.
 
Our current
approach allows for predicting the intensity dependence of shape
resonances in arbitrary diatomic molecules, based solely on their 
scattering length, $C_6$ coefficient and reduced mass, and on the 
polarizability of the constituent atoms, without knowledge of the full
interaction potential. This  is important for
utilizing non-resonant light control in molecule formation via
photoassociation~\cite{GonzalezPRA12} or Feshbach
resonances~\cite{TomzaPRL14} as it allows to predict the required
intensities. In addition to tuning the position and width of shape or
Feshbach resonances, non-resonant light control can also be employed
to change the background scattering length. This will be studied in
detail elsewhere~\cite{AM_SL}. 

\begin{acknowledgments}
Laboratoire Aim\'{e} Cotton is "Unit\'{e} Propre UPR 3321 du CNRS
associ\'{e}e \`{a} l'Universit\'{e} Paris-Sud", member of the
"F\'{e}d\'{e}ration Lumi\`{e}re Mati\`{e}re" (LUMAT, FR2764) and of
the "Institut Francilien de Recherche sur les Atomes Froids" (IFRAF).
R.G.F. gratefully acknowledges a Mildred Dresselhaus award from the
excellence cluster "The Hamburg Center for Ultrafast Imaging
Structure, Dynamics and Control of Matter at the Atomic Scale" of the
Deutsche Forschungsgemeinschaft and financial support by the Spanish
Ministry of Science FIS2011-24540 (MICINN), grants P11-FQM-7276 and
FQM-4643 (Junta de Andaluc\'{\i}a), and by the Andalusian research
group FQM-207. 
\end{acknowledgments}

\appendix
\section{Computational details}
\label{app:nodal}
In the following we present the computational details of the nodal
line technique  
applied to the asymptotic model for diatomics in a non-resonant field
for the tasks of determining the energy and wave
function of bound levels below the field shifted dissociation
limit, and the energy profile and properties of shape resonances.
All numerical calculations were performed using {\small{MATHEMATICA}}.  

\subsection{Bound levels}
\label{app:subsec:bound}
For a given value of energy, $e=-k^2$, $n$ linearly-independent solutions
${\bf y}^j$ are obtained by inward integration. Each solution $j$ is 
related to a specific channel $\ell_j=2(j-1)$ by imposing
${y}^j_\ell(x)$ to behave asymptotically as 
${y}^j_\ell(x)\propto \exp(-kx)$ in the $\ell_j$ channel
and zero in all others.
The physical solution ${\bf z}$ is a linear combination,
\begin{equation}
  \label{eq:lincomb}
  {\bf z}=\sum_{j=1}^n a_j{\bf y}^j\,, 
\end{equation}
where the radial component in
each channel $\ell$ must vanish on the corresponding nodal line
$x_{0\ell}$. The  resulting linear system of $n$ equations with $n$
unknown variables $a_j$ has a non-trivial solution if and only if 
\begin{equation}
  \label{eq:bound}
  D_{bound}(e)=\mathrm{det}\left(y^j_\ell(x_{0\ell})\right) = 0\,. 
\end{equation}
Equation~\eqref{eq:bound} is solved 
either by iteration on the energy $e$ or by interpolation of
$D_{bound}(e)$ on a set of $e$-values and finding the corresponding
zeros, $e_{bound}$. Solution of the linear system of the $n$ equations
$\sum_{j=1}^n a_j{\, y}^j_l(x_{0\ell})=0$ corresponding to the $n$ $\ell$-values
at an energy $e_{bound}$ yields the coefficients $a_j$ and thus 
the bound state wave functions of the various $\ell$-channels. The
coupled wave function ${\bf z}$ at the energy $e_{bound}$ is normalized to
one, such that hybridization can be measured by 
the weights $\varpi_\ell(e_b)=
\int_{x_{{0\ell}}}^{x_\infty}[\sum_j a_j
  y^j_{\ell}(x)]^2 \,dx$. 
	
\subsection{Resonance profiles} 
\label{app:subsec:profiles}
\begin{figure}[tb]
  \centering
  \includegraphics[width=0.9\linewidth]{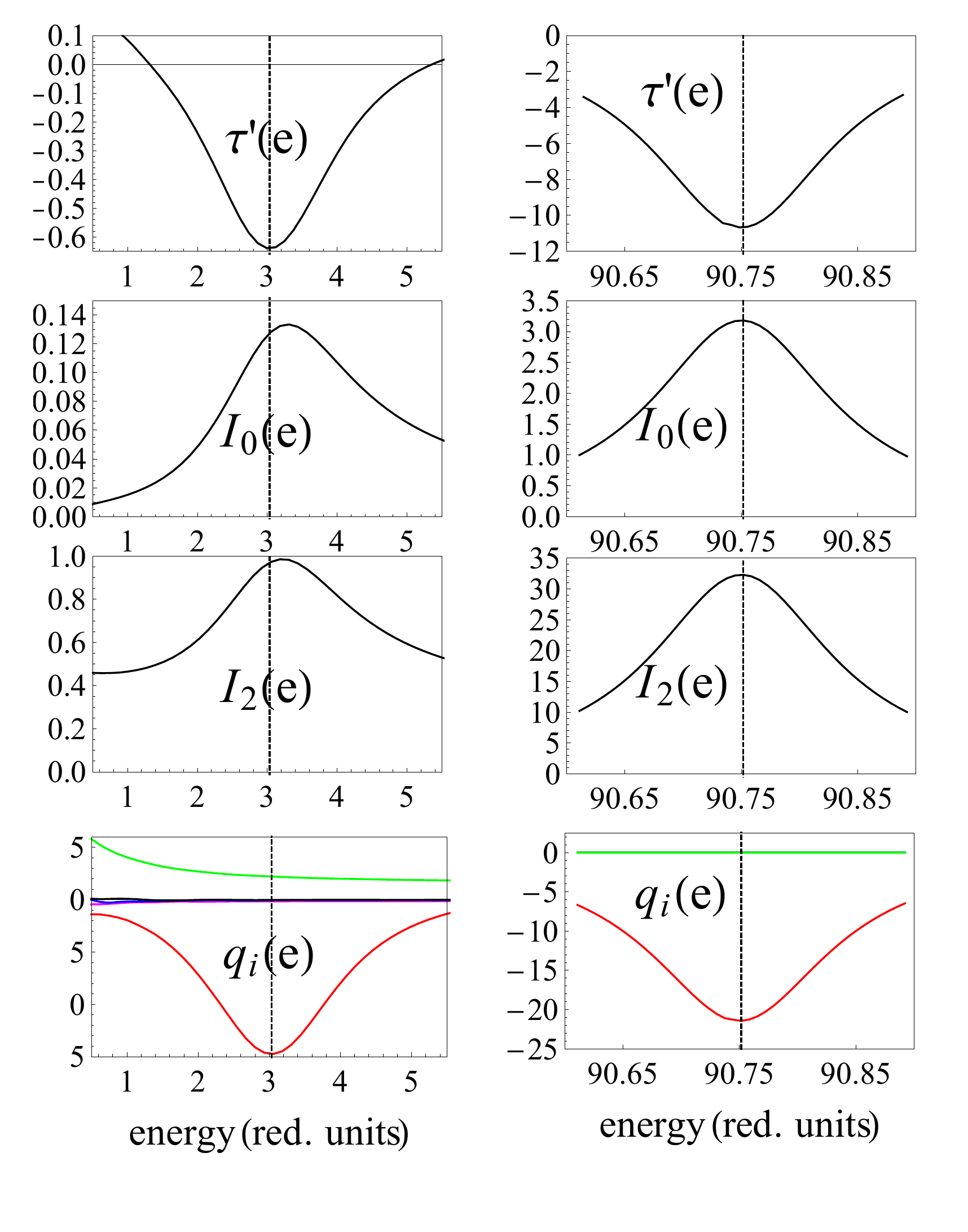}
  \caption{Comparison of four different methods allowing to characterize 
	a shape resonance, in two particular cases: in the left column, a very broad 
	resonance, ${\tilde \ell}=2$ of $^{86}$Sr$^{88}$Sr, at a field intensity 
	$i=5$~reduced units (calculated with 5 coupled channels), in the right one, 
	a narrow resonance, ${\tilde \ell}=8$ of $^{88}$Sr$_2$, at a field intensity 
	$i=6.5$~reduced units (calculated with 11 channels). In each column are 
	represented, from top to bottom, the energy variations of: the derivative 
	with respect to energy $\tau'(e)$ of the sum of the eigenphases of the K-matrix, 
	the total population density $I_0(e)$ trapped behind the centrifugal barriers, 
	the mean value $I_2(e)$ of the operator $1/x^2$ and the eigenvalues $q_i(e)$ 
	of the delay-matrix. One notices that all eigenvalues are zero (or positive) 
	except the lowest one (in red), which exhibits a negative Lorentzian 
	profile (indicating a positive time delay for the scattering). For a narrow 
	resonance (right column) all profiles are quite similar. For a broad resonance 
	(left column) the different methods yield slightly different profiles.} 
  \label{fig:profiles}
\end{figure}

To analyze the profiles of shape resonances in the $n$-fold degenerate
continuous spectrum, we use multichannel scattering
theory~\cite{seaton83,fano86}. For each energy $e$, $e=k^2>0$, we
calculate $2n$ particular, linearly-independent, energy-normalized 
solutions of the 2D Schr\"odinger equation~\eqref{eq:asyvect} by
inward integration. The initial conditions are taken at large 
distance $x_\infty$, where  the centrifugal term $1/x^2$ prevails. 
For each channel $\ell_j$ (where again $\ell_j=2(j-1)$), we determine
two particular solutions,  denoted by ${\bf j}^j(x)$ and
${\bf y}^j(x)$,  respectively, 
by imposing as asymptotic behavior in this channel either a regular
$\sqrt{(\pi x)/2}\,J_{\ell+1/2}(kx)$ or an irregular $\sqrt{(\pi
  x)/2}\,Y_{\ell+1/2}(kx)$ energy-normalized Bessel function and zero
in all other  channels.  
The physical solutions of Eq.~\eqref{eq:asyvect} are $n$ linear
combinations of the $2n$ calculated particular solutions which vanish 
which is related to the scattering matrix ${\bf S}$~\cite{seaton83,fano86} by 
on the nodal lines in each channel. Among all the possible sets of $n$
particular combinations, we choose the "standard" ones, ${\bf z}^j$,
which asymptotically contains a regular component  in the channel
$\ell_j$ only,
\begin{equation}
  \label{eq:Kfct}
  {\bf z}^j(x) = \sum_{j'=1}^n [\,\delta_{j',j}\,{\bf j}^{j'}(x)\, +\, 
  {\bf K}_{j'}^{j}\, {\bf y}^{j'}(x)]\,,
\end{equation}
where ${\bf K}$ is the so-called reaction matrix~\cite{seaton83,fano86}. 
Introducing two $n\times n$ matrices,
\begin{subequations}
  \label{eq:Mmatrices}
  \begin{eqnarray}
    \label{eq:Mmatrix}
    ({\bf M_{reg}})^j_\ell &=&  j^{j}_\ell(x_{0\ell})\,, \\
    ({\bf M_{irreg}})^j_\ell &=&  y^{j}_\ell(x_{0\ell})\,,
  \end{eqnarray}
\end{subequations}
the condition that the wave functions vanish on the nodal lines allow
us to determine the ${\bf K}$-matrix,
\begin{equation}
  \label{eq:Kmatrix}
  {\bf K} = - ({\bf M_{irreg}})^{-1}\cdot {\bf M_{reg}}\,,
\end{equation}
\begin{equation}
  \label{eq:Smatrix}
  {\bf S} =  ({\bf 1}+{\imath}{\bf K})\,\cdot\,({\bf 1}-{\imath}{\bf K})^{-1}\,.
\end{equation}
The existence and properties of a shape resonance can be determined by 
several different methods. An example of four different profiles that we 
have obtained in two particular cases ($^{86}$Sr$^{88}$Sr, ${\tilde \ell}=2$, 
at a field intensity $i=5$~reduced units and $^{88}$Sr$_2$, ${\tilde \ell}=8$, 
at a field intensity $i=6.5$~reduced units) is displayed in Fig.~\ref{fig:profiles}.

Studying the energy variations of the time-delay matrix 
${\bf Q}(e)$~\cite{fano86,smithPR60},
\begin{equation}
  \label{eq:Qmatrix}
  {\bf Q} = -\imath\, {\bf S}^\dagger \,\cdot\,\frac{d{\bf S}}{de}\,,
\end{equation}
and of its eigenvalues $q_j(e)$ and of the corresponding eigenvectors is probably 
the best adapted method. When a shape resonance is present, the wavepacket 
associated to an eigenvector is resonantly delayed during its scattering by 
the attractive potential. In the case of a narrow and isolated resonance, the 
lowest eigenvalue exhibits a negative Lorentzian profile,
\begin{equation}
  \label{eq:lorentzian}
  q_1(e) \sim -\frac{\gamma_r}{(e-e_r)^2+ (\frac{1}{2}\gamma_r)^2}\,,
\end{equation}
where $e_r$ is the resonance energy and $\gamma_r$ its FWHM. 
The  lifetime $T={\tau} t$  of the resonance (in SI units)
is calculated from the reduced lifetime  $t=1/\gamma_r$ and the
reduced unit of time $\tau=\hbar / \epsilon$.   
The channel-mixing of the resonance can be characterized by the
eigenvector corresponding to $q_1(e_r)$, which gives the partial wave 
decomposition of the continuum wave function (one among the $n$ wave functions 
of the $n-$multiple continuum) which concentrates the resonant character
of the scattering at this intensity value.

The second method consists in diagonalizing the ${\bf K}$-matrix, with
eigenvalues $\tan(\pi{\tau_j})$. The corresponding eigenvectors can be 
used to construct the so-called 'eigenchannel wave functions', ${\bf u}^{j}$, 
wich have the same asymptotic behavior, 
$u^j(x)\propto J_{\ell+1/2}(kx)\,+\tan(\pi{\tau_j})\,Y_{\ell+1/2}(kx)$, 
in all channels. The $\tau_j$ are called the eigenphase shifts. The total
eigenphase shift, 
\begin{equation}
  \label{eq:tau}
  \tau(e) =  \sum_{j=1}^n\,\tau^j\,,
\end{equation}
increases by $\pi$ when $e$ passes through the resonance
energy. The derivative with respect to the energy of the total 
eigenphase shift, $\tau'(e)$, exhibits a resonance profile, since  
it is related to the trace of the ${\bf Q}$ matrix by 
$\mathrm{Tr}\left[{\bf Q} (e)\right] = 2\tau'(e)$. 

It is finally possible to characterize the profile of a shape resonance
from  the radial components $v^j_\ell(x)$ of any orthonormalized set of
continuum wave functions. In practice, we have used energy-normalized wave
functions associated to eigenvalues of the matrix ${\bf K}^T\cdot{\bf K}$. 
Introduction of the transposed matrix ${\bf K}^T$ allows one to
eliminate numerical problems related to small asymmetries of the
matrix ${\bf K}$. In addition, the scalar product of the standard
functions ${\bf z}^j$ defined in Eq.~\eqref{eq:Kfct} is equal to
$\openone+ {\bf K}^T\cdot{\bf K}$. Since a shape resonance is a metastable state 
in which two atoms are temporally kept close to each other, a resonance profile 
is also expected for the density (per energy unit) inside the barrier and for the 
expectation value of the $1/x^2$. Precisely we calculate the following integrals
\begin{equation}
  \label{eq:x-2}
  I_p=\sum_{j=1}^n\,\sum_{\ell=1}^n \int_{x_{0\ell}}^{x^\ell_{max}}
  \,\frac{[z^j_\ell(x)]^2}{x^p}\,dx   \,.
\end{equation}
with either $p=0$ or $p=2$; $x^\ell_{max}$ is a very large value for $p=2$; 
for $p=0$, it is the position of the top of the centrifugal barrier 
${x^\ell_{top}}=[\ell(\ell+1)/3]^{-1/4}$ for $\ell>0$ and it is taken as
${x^2_{top}}$ for $\ell=0$.

As shown in Fig.~\ref{fig:profiles}, the four calculated profiles,
$q_1(e)$, $\tau'(e)$, $I_2(e)$ and $I_0(e)$,
exhibit similar shape, especially for narrow resonances. In the latter 
case, the profiles $q_1(e)$ are perfectly given by Eq.~\ref{eq:lorentzian}.

\subsection{Resonances via the complex energy method}  
\label{app:subsec:siegert}
Shape resonances with large $\ell$
are very narrow, even for low field intensity. This is due to the
presence of the broad and high potential barrier. Resonances lying
very close to the field-shifted dissociation 
limit also have a very small width. It is quite difficult to detect
narrow resonances and to calculate their characteristics from an 
analysis of the resonance profiles as described in
Appendix~\ref{app:subsec:profiles}. As an alternative, we therefore 
calculate the resonances as Siegert states
with complex energy $e_S=e -\imath\gamma/2$ where the
real and imaginary part are related to the resonance
energy and width~\cite{SiegertPR39,SimonsIJQC81}. Siegert states are
described by a 
complex wave function ${\bf z}$ whose asymptotic behavior
corresponds to an outgoing wave in each channel.  

To determine the Siegert states, we proceed similarly as for bound
levels, cf. Appendix~\ref{app:subsec:bound}. We first determine $n$
linearly-independent particular complex solutions ${\bf y}^j$ of
Eq.~\eqref{eq:asyvect}. Changing $e$ into $e_S=k^2$ results in a
complex $k$-value. Inward integration, imposing an outgoing wave
asymptotically in the $\ell_j$ channel and zero in all others
yields the ${\bf y}^j$. The outgoing wave is written
as a combination of the regular and 
irregular Bessel functions, $(-1)^{(\ell+1)/2} \sqrt{\pi x/2} 
\,\left[J_{\ell +1/2}(kx)+\imath Y_{\ell +1/2}(kx)\right]$.
The physical Siegert wave function, obtained as a linear combination of
the ${\bf y}^j$ solutions, has to satisfy the boundary condition at
small $x$, {\it i.e.}, the radial components in all channels have to
vanish on the corresponding nodal line $x_{0\ell}$, calculated at the
energy $\mathfrak{Re}(e_S)$. These conditions are equivalent to a
vanishing determinant $D_{siegert}(e_S)$ of the radial
$\ell$-components of the $n$ particular solutions $ y^j_\ell(x)$ at
the node positions. This condition quantifies the resonance energy to
the value $e_{S,r}=e_r -\imath\gamma_r/2$. 


\end{document}